\documentclass{PoS}
\newcommand{\beq}{\begin{equation}}
\newcommand{\eeq}{\end{equation}}
\newcommand{\cA}{{\cal A}}
\newcommand{\cAb}{{\overline{\cal A}}}
\newcommand{\cF}{{\cal F}}
\newcommand{\cFb}{{\overline{\cal F}}}
\newcommand{\cD}{{\cal D}}

\newcommand{\cDb}{{\overline{\cal D}}}
\newcommand{\cU}{{\cal U}}
\newcommand{\cUb}{{\overline{\cal U}}} 
\newcommand{\KD}{{K\"{a}hler-Dirac }}
\newcommand{\cN}{{\cal N}}
\newcommand{\Tr}{{\rm Tr\;}}
\newcommand{\bx}{{\bf x}}

\title{Three dimensional lattice gravity as supersymmetric Yang-Mills theory }

\ShortTitle{Topological lattice gravity}

\author{\speaker{Simon Catterall}\thanks{Supported in part by DOE grant DE-FG02-85ER40237}\\
        Department of Physics, Syracuse University, Syracuse NY 13244\\
        E-mail: \email{smcatterall@gmail.com}}

\abstract{We argue that a certain twisted supersymmetric Yang-Mills theory in three
dimensions with gauge group $SU(2)$ possesses a set of topological observables
whose expectation values can be computed in a related 
Chern Simons theory. This Chern Simons theory has been proposed as
a definition of three dimensional
Euclidean quantum gravity. Since the YM theory admits a discretization which
preserves the values of topological observables we conjecture
that it can be used as
a non-perturbative definition of the quantum gravity theory.} 

\FullConference{The XXVIII International Symposium on Lattice Field Theory, Lattice2010\\
		June 14-19, 2010\\
		Villasimius, Italy}

\begin{document}

\section{Introduction}
In recent years there has been a resurgence of interest in the problem of
formulating supersymmetric
lattice theories. This has largely been driven by
the realization that in certain classes of theory a discretization
can be derived which preserves a subset of
the continuum supersymmetry - see for example the recent review \cite{Catterall:2009it} and references therein. 

For our purposes in this paper the key idea that lies behind these new
constructions is that of {\it topological twisting} -- one discretizes not the original
supersymmetric theory but a theory formulated in terms a new set of {\it twisted} variables.
This twisting procedure has the merit of exposing a nilpotent, scalar supercharge which is
compatible with an underlying lattice structure - the associated fermionic symmetry holds
exactly on the lattice and leads to a number of remarkable properties; the vacuum
energy is zero to all orders in the lattice spacing, the boson-fermion spectrum
is degenerate and the lattice
theory contains cut-off independent observables whose expectation values can be
computed exactly and which are independent of the coupling constant. These
latter observables correspond to a topological sector of the continuum theory; a set of
observables whose expectation values are independent of smooth deformations in 
the background space. This independence on the background metric is also common to
certain Chern Simons theories and specifically can be used to
write down background independent formulations of topological quantum gravity \cite{Witten:1988hc, Witten:1989sx,Chamseddine:1989nu}.

The three dimensional Chern Simons gravitational theory has been well studied and has been
shown to correspond to
the usual Einstein-Hilbert theory \cite{Carlip:1995zj}. In this talk we show that
the relevant gravitational observables appearing in this three dimensional Chern
Simons theory can be computed in an associated twisted Yang-Mills theory in three
dimensions. Furthermore, the lattice theory that targets this continuum
twisted theory can be discretized and hence can serve as a non-perturbative definition of
the quantum gravity theory.

In this talk we will first review the Chern Simons formulation of (Euclidean) quantum gravity and
then go on to show that the relevant topological observables can be computed in
an equivalent continuum twisted Yang-Mills theory with
$\cN=4$ supersymmetry. We then go to review the construction of
a lattice theory which targets this twisted YM theory in the naive continuum limit.
 
\section{Chern-Simons formulation of three dimensional gravity}

It has been known for a long time that
three dimensional gravity can be reformulated in the language of
gauge theory \cite{Witten:1988hc,Witten:1989sx}. 
The construction employs
a Chern-Simons action and in Euclidean space the relevant local symmetry
corresponds to the group $SO(3,1)\sim SL(2,C)$ - the complexified
$SU(2)$ group. Notice that this includes both local Lorentz rotations and
translations. Furthermore the resulting theory is topological
and global gravitational observables are
determined by integrals over the moduli space of flat $SL(2,C)$ connections. We show
these properties explicitly below.

Consider the following three dimensional Chern-Simons action
\beq
\int d^3 x\epsilon_{\mu\nu\lambda}{\rm \hat{Tr}}\left(A_\mu F_{\nu\lambda}
-\frac{1}{3}A_\mu \left[A_\nu , A_\lambda\right]\right)
\label{CS}\eeq
Furthermore, assume that the gauge field $A_\mu$ takes values
in the adjoint representation of the group $SO(3,1)$. A convenient
representation for the six generators of the Lie algebra of
this group is then given by commutators
of the (Lorentzian) Dirac matrices 
$\gamma^{AB}=\frac{1}{4}\left[\gamma^A,\gamma^B\right]$ where
$\left(\gamma^a\right)^\dagger=\gamma^a\;a=1\ldots 3$ and
$\left(\gamma^4\right)^\dagger=-\gamma^4$
This yields an expression for the gauge field of the form
\beq
A_\mu=\sum_{A<B}A_\mu^{AB}\gamma^{AB}\quad A,B=1\ldots 4\eeq
Finally, the group indices are contracted using the invariant tensor
$\epsilon_{ABCD}$ corresponding to a trace of the form
\beq
{\rm \hat{Tr}}(X)={\rm Tr}(\gamma_5 X)\eeq

To see explicitly that the resulting theory is just three dimensional
gravity we decompose the gauge field and field strength in terms of
an $SO(3)$ subgroup
\begin{eqnarray}
A_\mu&=&\sum_{a<b}\omega_\mu^{ab}\gamma^{ab}+\frac{1}{l}e_\mu^{a}\gamma^{4a}\quad
a,b=1\ldots 3\nonumber\\ 
F_{\mu\nu}^{ab}&=&\sum_{a<b} \left(R_{\mu\nu}^{ab}+\frac{1}{l^2}e_{\left[\mu\right.}^a
e_{\left.\nu\right]}^b\right)\gamma^{ab}\nonumber\\
F_{\mu\nu}^a&=&\sum_a D_{\left[\mu\right.}e_{\left.\nu\right]}^a
\end{eqnarray}
The covariant derivative appearing in the field strength contains
just the $SO(3)$ gauge field $\omega_\mu$ and we have
introduced a explicit length scale $l$ into the definition of
the gauge fields $e_\mu$.
After substituting into the Chern-Simons
action one recognizes that it corresponds to 
three dimensional Einstein-Hilbert gravity including
a cosmological constant and written in the
first order tetrad-Palatini formalism \cite{Witten:1988hc}.
\beq
S_{\rm EH}=\frac{1}{l}\int \epsilon^{\mu\nu\lambda} \epsilon_{abc}\left(
e^a_\mu R^{bc}_{\mu\nu}-\frac{1}{3l^2}e^a_\mu e^b_\nu e^c_\lambda\right)
\label{palatini}\eeq
with $\omega_\mu$ and $e_\mu$ corresponding to the spin connection
and dreibein and $1/l^2$ playing the role of a cosmological
constant.

To see that this theory is classically equivalent to the usual metric
theory of gravity one merely has to notice that the
classical equations of the Chern-Simons theory require that 
the $SO(3,1)$ field strength vanish $F_{\mu\nu}^{AB}=0$. One consequence of
this is that the torsion $T=F^{4a}=D_{\left[\mu \right.} e_{\left. \nu\right]}$ must
vanish.
This condition allows one to solve for the spin connection as a function of the
dreibein $\omega=\omega(e)$ and ensures that the curvature appearing in eqn.~\ref{palatini}
is indeed nothing more than the usual Riemann curvature associated to a torsion free
connection determined by the metric $g_{\mu\nu}=e^a_\mu e^a_\nu$.

Indeed the condition $F_{\mu\nu}=0$ also requires that 
the $SO(3)$ curvature $R_{\mu\nu}$ must take the constant value
$-1/l^2$. Such a solution corresponds (at least locally)
to hyperbolic three space $H^3\sim SO(3,1)/SO(3)$ which is the correct
solution to a metric theory of three dimensional
Euclidean quantum gravity in the presence of a cosmological
constant.

Finally one can show that the theory restricted to
this space of flat connections is also invariant under
diffeomorphisms \cite{Witten:1988hc} at least on shell. 
This result follows from the fact that one can express
a general coordinate transformation on $A_\mu$
with parameter $-\xi^\nu$ as
a gauge transformation with parameter $\xi^\mu A_\mu$ plus a term
which vanishes on flat connections.
\beq
\delta A_\mu^\xi=-D_\mu (\xi^\nu A_\nu)-\xi^\nu F_{\mu\nu}\eeq

The classical solutions of this Chern Simons theory thus correspond to the space
of flat $SL(2,C)$ (or equivalently $SO(3,1)$) connections
up to $SL(2,C)$ gauge transformations. One might ask how such a situation
can correspond to
gravity since this solution does not seem to allow for the propagation of
any local degrees of freedom. In fact, it is well known that three
dimensional gravity possesses no gravitons 
so this result should not be too surprising \cite{Carlip:1995zj}. 

However, notice that there may still
be non-local observables in such a theory corresponding to the
expectation value of Wilson lines of the $SL(2,C)$ group which cannot be shrunk to a point.  

Since the expectation values of such Wilson lines cannot depend on any
background metric introduced to construct the Chern Simons theory they
are explicitly topological in nature. It is
important to stress that {\it physical spacetime} is not determined by the choice of 
background metric but instead determined dynamically by the values of the dreibeins. 

We now turn to a different theory based on a twisted supersymmetric Yang-Mills
which will turn out to possess the same
vacuum state and the same set of topological observables whose vacuum
expectation values will also be background independent.

\section{Twisted $\cN=4$ gauge theory in three dimensions}

The twist of $\cN=4$ super Yang-Mills that we are interested in can be 
most succinctly written in the form 
where 
\begin{eqnarray}
S &=&\frac{1}{g^2}Q\int d^3x\;\sqrt{h} \left(\chi^{\mu\nu}\cF_{\mu\nu}+
\eta\left[\cDb^\mu,\cD_\mu\right]+\frac{1}{2}\eta d+B_{\mu\nu\lambda}\cDb^\lambda\chi^{\mu\nu}\right)\nonumber\\
\label{action1}
\end{eqnarray}
The fermions comprise a multiplet of p-form fields 
$(\eta,\psi_\mu,\chi_{\mu\nu},\theta_{\mu\nu\lambda})$\footnote{It is common in the
continuum literature to replace the 2 and 3 form fields in these
expressions by
their Hodge duals; a second vector $\hat{\psi}_\mu$ and scalar $\hat{\eta}$ see,
for example \cite{Blau:1996bx}} where in three
dimensions $p=0\ldots 3$. This multiplet of twisted
fermions corresponds to a single \KD field and here possesses eight single
component fields as expected for a theory with $\cN=4$ supersymmetry in
three dimensions.

The imaginary parts of the
complex gauge field $\cA_\mu\;\mu=1\ldots 3$ appearing in this construction
yield the three scalar fields of the conventional (untwisted) theory. Fields $d$ and
$B_{\mu\nu\lambda}$ are auxiliaries introduced to render the scalar
nilpotent supersymmetry $Q$ nilpotent off shell.
The latter acts on the twisted fields as follows
\begin{eqnarray}
Q \cA_\mu&=&\psi_\mu\nonumber\\
Q \cAb^\mu&=&0\nonumber\\
Q \psi_\mu&=&0\nonumber\\
Q \chi^{\mu\nu}&=&\cFb^{\mu\nu}\nonumber\\
Q \eta&=&d\nonumber\\
Q d&=&0\nonumber\\
Q B_{\mu\nu\lambda}&=&\theta_{\mu\nu\lambda}\\
Q \theta_{\mu\nu\lambda}&=&0
\label{susy}
\end{eqnarray}
where the background metric $h_{\mu\nu}$ is used to raise and
lower indices in the usual manner
$\chi^{\mu\nu}=h^{\mu\alpha}h^{\nu\beta}\chi_{\alpha\beta}$
and is a $Q$-singlet. The topological character of the
theory follows from the $Q$-exact structure of $S$\footnote{Notice that this
construction differs slightly from the one discussed in \cite{Catterall:2010} in that one of the
fermion terms is here trivially rewritten as a $Q$-exact rather than
$Q$-closed form. This does not affect any of the subsequent arguments} 

The complex covariant derivatives appearing in these expressions are defined
by
\begin{eqnarray}
\cD_\mu&=&\partial_\mu+\cA_\mu=\partial_\mu+A_\mu+iB_\mu\nonumber\\
\cDb^\mu&=&\partial^\mu+\cAb^\mu=\partial^\mu+A^\mu-iB^\mu 
\end{eqnarray}
while all fields take values in the adjoint
representation of $SU(N)$\footnote{The generators are taken to be {\it
anti-hermitian} matrices satisfying $\Tr (T^aT^b)=-\delta^{ab}$}.
It should be noted that despite the appearance of a complexified
connection and field strength the theory possesses only the usual
$SU(N)$ gauge invariance corresponding to the real part of the gauge field.
In our current application we will
need to consider only the case where the group is $SU(2)$ although the
topological character of this theory holds for any value of $N$.

The structure of this
twisted theory is similar to that of the Marcus twist of
$\cN=4$ super Yang-Mills in four dimensions \cite{Marcus,Unsal:2006qp,
Catterall:2007kn} 
which plays an important role in the Geometric-Langlands 
program \cite{Kapustin:2006pk}. 

Doing the $Q$-variation, integrating out the field $d$ and using the
Bianchi identity \beq
\epsilon_{\mu\nu\lambda}\cDb^\lambda \cFb^{\mu\nu}=0\eeq
yields
\beq
S=\frac{1}{g^2}\int d^3 x\;\sqrt{h}\left(L_1+L_2\right)\eeq
where
\begin{eqnarray}
L_1&=&\Tr \left(
-\cFb^{\mu\nu}\cF_{\mu\nu}+\frac{1}{2}[ \cDb^\mu, \cD_\mu]^2)\right)\nonumber\\
L_2&=&\Tr \left(
-\chi^{\mu\nu}\cD_{\left[\mu\right.}\psi_{\left.\nu\right]}-
\psi_\mu\cDb^\mu\eta -\theta_{\mu\nu\lambda}\cDb^{\left[\lambda\right.}
\chi^{\left.\mu\nu\right]}\right)
\label{action}
\end{eqnarray}
The terms appearing in $L_1$ can then be written
\begin{eqnarray}
\cFb^{\mu\nu}\cF_{\mu\nu}&=&(F_{\mu\nu}-[B_\mu,B_\nu])
                            (F^{\mu\nu}-[B^\mu,B^\nu])+
(D_{\left[\mu\right.}B_{\left.\nu\right]})
(D^{\left[\mu\right.}B^{\left.\nu\right]})\nonumber\\
\frac{1}{2}\left[\cDb^\mu,\cD_\mu\right]^2 &=& -2\left(D^\mu B_\mu\right)^2
\end{eqnarray}
where $F_{\mu\nu}$ and $D_\mu$ denote the usual field strength and
covariant derivative depending on the real part of the connection $A_\mu$.
The classical vacua of this theory correspond to solutions of
the equations
\begin{eqnarray}
F_{\mu\nu}-[B_\mu,B_\nu]&=&0\nonumber\\
D_{\left[\mu\right.}B_{\left.\nu\right]}&=&0\nonumber\\
D^\mu B_\mu&=&0
\label{moduli}
\end{eqnarray}
The same moduli space arises in
the study of the Marcus twist of four dimensional
$\cN=4$ Yang-Mills where it is argued to correspond to
the space of flat {\it complexified} connections modulo {\it complex} gauge
transformations. A simple way to understand this is to recognize
that the additional term $D^\mu B_\mu=0$ appearing in the vacuum
equations~\ref{moduli} resembles a partial gauge fixing   
of a theory with a complexified gauge invariance 
and associated gauge fields $A_\mu$ and $B_\mu$
down to a theory
possessing just the usual $SU(N)$ (here $N=2$) 
gauge invariance implemented via the gauge field $A_\mu$.

More specifically,
Marcus showed in \cite{Marcus} that the solutions of eqns.~\ref{moduli}
modulo $SU(2)$ gauge transformations are in one to one
correspondence with the space of
flat complexified $SU(2)=SL(2,C)$ connections modulo complexified gauge
transformations. These arguments should hold in
the three dimensional case too.

The topological character of the theory
then guarantees that any $Q$-invariant observable such as the
partition function can be evaluated {\it exactly} by considering only
Gaussian fluctuations about such vacuum configurations. Furthermore, 
it is easy to see from eqn.~\ref{action1}
that the energy momentum tensor of this
theory is $Q$-exact rendering the expectation values of such
topological observables independent of smooth deformations
of the background metric $h_{\mu\nu}(x)$.

Most importantly notice that the fact that the complexified
connection $\cAb_\mu$ is a $Q$-singlet
allows us to trivially construct a class of topological observables
corresponding to the trace of an associated Wilson loop.
\beq 
O(\gamma)={\cal P} e^{ \int_\gamma \cAb^\mu .dx_\mu}\eeq 
If this loop
is non-contractible (for example if
it winds around a cycle of a torus)
we obtain the same non-trivial global topological observable we encountered in
the Chern-Simons construction of three dimensional Euclidean gravity.

We can cement this connection between
the twisted Yang-Mills theory and the gravitational
theory by identifying the imaginary parts
of the $SL(2,C)$ connection - the field $B_\mu$ occurring
in the Yang-Mills theory - with the
matrix valued field
$e_\mu$ occurring the tetrad-Palatini action. 
Notice that the fact that
the field $B_\mu$ transforms in the adjoint representation
of the $SU(2)$ gauge group translates in the
gravitational theory to the statement that
the dreibein $e_\mu^a$ transforms 
as a {\it vector} under local Lorentz
transformations just as it should.

These considerations together with the
equivalence of the topological sectors
of these two theories leads us
to conjecture that the twisted two color Yang-Mills gives an
alternative representation of the gravity theory. 
Furthermore, this alternative representation has some
advantages -- the path integral is now well defined and indeed
may be given a non-perturbative definition as the appropriate
limit of a gauge and supersymmetric invariant lattice model to which we now turn.

\section{Lattice theory}

The twisted theory described in the previous section may be discretized
using the techniques developed in \cite{Catterall:2007kn,Damgaard:2008pa,Damgaard:2007be}. 
The resultant lattice theories
have the merit of preserving both gauge invariance and the scalar
component of the twisted
supersymmetry. 
Here we show how to derive this lattice theory by direct discretization
of the continuum twisted theory. We will start by assuming
that the continuum theory
is formulated in flat (Euclidean) space 
with metric $h_{\mu\nu}=\delta_{\mu\nu}$. 

In the case of topological observables the choice of
metric is unimportant and hence the lattice theory we construct will
yield expectation values for topological operators which depend
only on the topology of the lattice and not on the coupling,
lattice spacing or the fact that we started by discretization of
a theory in a flat background. 

The transition to the lattice from the continuum theory 
requires a number of steps. The first, and most important,
is to replace the continuum complex gauge field $\cA_\mu(x)$ at every
point
by an appropriate complexified Wilson
link $\cU_\mu(\bx)=e^{\cA_\mu(\bx)},\mu=1\ldots 3$. 
These lattice fields 
are taken to be
associated with unit length vectors in the coordinate
directions $\mu$ in an abstract
three dimensional hypercubic lattice. 
By supersymmetry the fermion
fields $\psi_\mu(\bx),\mu=1\ldots 3$ lie on the same oriented
link as their
bosonic superpartners running from $\bx\to\bx+\mu$. In contrast the scalar
fermion $\eta(\bx)$ is associated with the site $\bx$
of the lattice and the
tensor fermions $\chi^{\mu\nu}(\bx),\mu<\nu =1\ldots 3$ 
with a set of diagonal face links
running from $\bx+\mu+\mu\to \bx$. The final 3 form field
$\theta_{\mu\nu\lambda}(\bx)$ is then naturally placed on the
body diagonal running from $\bx\to \bx+\mu+\mu+\lambda$. 
The construction then posits that
all link fields transform as bifundamental
fields under gauge transformations
\begin{eqnarray}
\eta(\bx)&\to& G(\bx)\eta(\bx) G^\dagger(\bx)\nonumber\\
\psi_\mu(\bx)&\to& G(\bx)\psi_\mu(\bx) G(\bx+\mu)\nonumber\\
\chi^{\mu\nu}(\bx)&\to&G(\bx+\mu+\mu)\chi^{\mu\nu}(\bx)G^\dagger(\bx)\nonumber\\
\cU_\mu(\bx)&\to&G(\bx)\cU_\mu(\bx)G^\dagger(\bx+\mu)\nonumber\\
\cUb^\mu(\bx)&\to&G(\bx+\mu)\cUb^\mu(\bx)G^\dagger(\bx)
\label{gaugetrans}
\end{eqnarray}
Notice that
we can keep track of the orientation of the lattice field by following
its continuum index structure -- upper index fields are placed on
negatively orientated links, lower index fields live on positively
oriented links. 

The action of the scalar supersymmetry on these fields is given by the
continuum expression in eqn.~\ref{susy} with the one modification
that the continuum field $\cA_\mu(x)$ is replaced with the
Wilson link $\cU_\mu(x)$ and the lattice field strength being defined
as $\cF_{\mu\nu}=\cD^{(+)}_\mu U_\nu$.
The supersymmetric and gauge invariant
lattice action which corresponds to
eqn.~\ref{action} then takes a very similar form
to its continuum counterpart
\begin{eqnarray}
S_1&=&Q\sum_{\bx} \left(\chi^{\mu\nu}\cF_{\mu\nu}+
\eta \left[ \cDb^{(-)\mu}\cU_\mu \right]+\frac{1}{2}\eta d\right)\nonumber\\
S_2&=&\sum_{\bx} \theta_{\mu\nu\lambda}\cDb^{(+)\lambda} \chi^{\mu\nu}
\label{act}
\end{eqnarray}
The covariant difference operators appearing in these expressions are
defined by 
\begin{eqnarray}
\cD^{(+)}_\mu f_\nu(\bx)&=&
\cU_\mu(\bx)f_\nu(\bx+\mu)-f_\nu(\bx)\cU_\mu(\bx+\mu)\nonumber\\
\cDb^{(-)\mu} f_\mu(\bx)&=&
f_\mu(\bx)\cUb^\mu(\bx)-\cUb^\mu(\bx-\mu)f_\mu(\bx-\mu)
\end{eqnarray}
These expressions are determined by the twin requirements that they reduce
to the corresponding continuum results for the adjoint covariant derivative
in the naive continuum
limit $\cU_\mu\to 1+\cA_\mu$ and that they transform
under gauge transformations like the corresponding lattice
link field carrying the same indices. This allows the terms in the
action to correspond to gauge invariant closed loops on the lattice.
The action can also be
shown to be free of fermion doubling problems -- see the
discussion in \cite{Catterall:2007kn}.

As in the continuum,
the presence of an exact $Q$-symmetry allows the definition of a 
class of supersymmetric Wilson loop corresponding to the trace of
the product of $\cUb_\mu$ links around a closed loop in the lattice. 
\beq
O=\prod_{t=1}^T \cUb^t(\bx)\eeq
In principle other
non-contractible loops can also be defined corresponding to knot-like structures as
in the continuum.
The vacuum expectation value of these operators can be computed
exactly by restriction to the moduli space of theory and can
probe only topological features of the theory. 

To summarize; we have constructed a supersymmetric lattice theory based on the group
$SU(2)$ which possesses a topological subsector which can be identified with
the set of global gravitational observables in a Chern-Simons reformulation of 3d
Euclidean quantum gravity. The lattice theory is well defined and we conjecture that
the lattice theory may
constitute a non-perturbative definition of the gravitational theory. It is an open
question as to whether any similar correspondences between Yang-Mills theories and
topological Chern Simons theories can be made in higher dimensions. The author would
like to thank Poul Damgaard for useful discussions.

\end{document}